\documentclass[reprint,jcp,numerical,aip]{revtex4-1}

\usepackage[psamsfonts]{amsfonts}
\usepackage{amsmath}

\usepackage{color}

\def\mat#1{{\bf #1}} 
\newcommand \dd[1]  { \,\textrm d{#1}                       }

\newcommand{\be}{\begin{equation}}
\newcommand{\ee}{\end{equation}}

\begin{document}

\title{On Factorization of Molecular Wavefunctions}
\thanks{Manuscript published October 13, 2015: J. Phys. A: Math. Theor {\bf 48} (2015) 445201 (20 pages) }
\author {Thierry Jecko$^{1,}$\footnote{Electronic mail: thierry.jecko@u-cergy.fr}, Brian T. Sutcliffe$^{2,}$\footnote{Electronic mail: bsutclif@ulb.ac.be} and R. Guy Woolley$^{3}$\\
$^{1}$\emph{AGM, UMR 8088 du CNRS, Universit\'{e} de Cergy-Pontoise,
D\'{e}partement de math\'{e}matiques, site de Saint Martin,
2 avenue Adolphe Chauvin, F-95000 Pontoise, France. }\\
$^{2}$\emph{Service de Chimie quantique et Photophysique,  Universit\'{e} Libre de Bruxelles, B-1050 Bruxelles, Belgium} \\$^{3}$\emph{School of Science and Technology, Nottingham Trent University, Nottingham NG11 8NS, U.K.} }
\date{\today}

\begin{abstract}
Recently there has been a renewed interest in the chemical physics literature of factorization of the position representation eigenfunctions \{$\Phi$\} of the molecular Schr\"{o}dinger equation as originally proposed by Hunter in the 1970s. The idea is to represent $\Phi$ in the form $\varphi\chi$ where $\chi$ is \textit{purely} a function of the nuclear coordinates, while $\varphi$ must depend on both electron and nuclear position variables in the problem. This is a generalization of the approximate factorization originally proposed by Born and Oppenheimer, the hope being that an `exact' representation of $\Phi$ can be achieved in this form with
$\varphi$ and $\chi$ interpretable as `electronic' and `nuclear'
wavefunctions respectively. We offer a mathematical analysis of these proposals 
that identifies ambiguities stemming mainly from the singularities in the 
Coulomb potential energy. 
\end{abstract}

\maketitle
\section{Introduction}

In the Born-Oppenheimer (BO) model an eigenfunction $\Phi$ of the molecular Hamiltonian is often approximated by a simple product of `electronic' and `nuclear' wavefunctions
\begin{equation}
\Phi(r,R) \approx \varphi(r,R)\chi(R),~~~\langle\varphi|\varphi\rangle_{r} = 1~\mbox{for all}~R
\label{BOfn}
\end{equation}
where $\varphi$ is an eigenfunction of the clamped-nuclei electronic Hamiltonian, and $\chi$ is a vibration-rotation wavefunction for the nuclear motion on the potential energy surface (PES) associated with $\varphi$. $r$ and $R$ stand collectively for the electronic and nuclear coordinates respectively. This is probably the most important approximation in the quantum theory of molecules, and it has been subject to exhaustive mathematical analysis\cite{TJ:14}. It should be noted that the BO approximation is a singular perturbation problem based on the limit of infinite nuclear mass\cite{TJ:14, JMC:77, JMC:80}, and it is plausible that the product form, when valid, is a direct consequence of the limit. Nevertheless there has recently been renewed interest\cite{AMG:10, AMG:12, GG:14, LSC:13, LSC:14} in the question as to whether the $\approx$ symbol in (\ref{BOfn}) can be replaced by $=$ with suitably redefined `electronic' and `nuclear' functions $\overline\varphi$ and $\overline\chi$, a so-called `exact' factorization of an eigenfunction $\Phi$. The present work offers a mathematical analysis of this factorization.

The earliest attempt that we know of to write an exact wavefunction $\Phi(r,R)$
of the Coulomb Hamiltonian $\mathsf{H}$ for a neutral system of electrons and nuclei in a factorized form was made by Hunter\cite{H:75}. His analysis was made in terms of conditional and marginal probability amplitudes; the wavefunction $\Phi$, assumed normalized, is written in the product form
\be
\Phi(r,R)=f(R)\phi(r,R) 
\label{hunxct}
\ee
with the nuclear function $f(R)$ defined as a marginal by means of
\be
|f(R)|^2=\int\Phi(r,R)^*\Phi(r,R)\dd r.
\label{hunf}
\ee
This fixes $f(R)$ to within a phase factor,
\begin{equation}
e^{i\theta(R)}
\label{phase}
\end{equation}
where $\theta(R)$ is a real function of the nuclear coordinates. In the absence of a criterion to choose it, Hunter suggested taking simply
\begin{equation}
f(R)~=~-|f(R)|~\mbox{or}~f(R)=|f(R)|.
\label{sign}
\end{equation}

The associated function, $\phi$, is then defined to be the quotient
\be
\phi(r,R)=\frac{\Phi(r,R)}{f(R)} 
\label{hunewf}
\ee
and it satisfies the normalization condition
\[
\int\phi(r,R)^*\phi(r,R)\dd r = 1
\]
for all $R$. Obviously this construction would be problematic if either $f(R)$ has zeroes for finite $R$ or $\Phi(r,R)/f(R)$ is too 
irregular at infinity. The construction, if applicable, is available for any wavefunction $\Phi_{i}$\ and so the nuclear functions \{$f(R)_{i}$\} are required generally to be quite different from the usual approximate nuclear wavefunctions for vibrationally excited states which do have nodes \cite{GH:81}. Equally, it is evident that every wavefunction $\Phi_{i}$ has its own distinct electronic factor, $\phi_{i}$; this is to be contrasted with the BO description (\ref{BOfn}) where whole groups of approximate eigenfunctions \{$\Phi_\alpha$\} share one electronic state $\varphi$ which supports a vibration-rotation manifold of states.

More recently, attempts at factorization have been made by Gross and co-workers\cite{AMG:10, AMG:12, GG:14} using both time-independent and time-dependent formulations which are a development of Hunter's approach and, reverting to the time-independent form, Cederbaum has proposed a related factorization scheme\cite{LSC:13} (see also the subsequent Erratum\cite{LSC:14}); it is these more recent proposals we analyse here though we point out the connection between Hunter's early work\cite{H:75} and the more recent work of Gross \textit{et al.}. 

For a freely moving system it is always possible to separate completely the centre-of-mass dynamics (free motion) from the internal motions of 
the molecule. It is not essential for an account of factorization to remove
the centre-of-mass motion (and to do so complicates the form of the internal
Hamiltonian somewhat - see below) but if one does not, the description of the
bound-states of the molecule is more involved.
If we denote the position coordinate of the centre-of-mass by ${\bf R}$, and introduce a set of independent internal position variables, straightforward calculation\cite{SW:05} yields the Hamiltonian $\mathsf{H}$ in Schr\"{o}dinger representation separated into internal and centre-of-mass contributions
\begin{equation}
\mathsf{
H}= \mathsf{H}' - \frac{\hbar^2}{2M_T}\nabla^2_{\bf R}\ ,\ \mathsf{H}'=\mathsf{H}_{el} + \mathsf{T}_{n}.
\label{Hint}
\end{equation}
Here $M_T$ is the total molecular mass and $\mathsf{H}_{el}$, which only differentiates in the electronic variables, and $\mathsf{T}_n$ are defined explicitly in Section~\ref{CouHam}. The spectrum of the Hamiltonian 
$\mathsf{H}$ is purely continuous and the description of the molecular 
bound-states requires a rather complicated mathematical formulation. Indeed, 
$\mathsf{H}$ has no eigenfunctions. On the 
other hand the eigenfunctions of the internal molecular Hamiltonian $\mathsf{H}'$, 
(\ref{ellip}), are square integrable and have (distributional) derivatives up to second-order that are also square integrable; they belong to the Sobolev space ${\cal{H}}^2$. These are the true bound-states 
\{$\Psi_i$\} associated  with discrete energies \{$E_i$\} that describe the 
internal motions of the molecule. The regularity properties of the bound-state wavefunctions turn out to be important for a 
precise account of factorization, and it is advantageous to frame the discussion directly in terms of the internal Hamiltonian.

The paper is organized as follows; in Section~\ref{CouHam} we review some features of the Schr\"{o}dinger equation for the Coulomb Hamiltonian which are pertinent here.  
We emphasize the occurrence of singularities in the Coulomb potential energy operator which require that the molecular Schr\"{o}dinger equation (in position representation) be interpreted in a more general setting than a classical partial differential equation. Factorization of an eigenfunction of the
Coulomb Hamiltonian evidently must take account of this mathematical setting,
but it also brings in new problems which this paper aims to characterize. In the
following we shall identify some ambiguities in the recent discussions of
factorization and describe a mathematical framework where the ambiguities are
removed. Some remaining uncertainties are reported. A common feature of the proposed factorization schemes is that the factors ($\overline{\varphi}, \overline{\chi}$) are solutions of a system of non-linear equations. In Section~\ref{s:cederbaum-1} we study the formal computation that gives rise to these equations, paying close attention to the regularity properties required to give it a precise meaning. In Section~\ref{s:cederbaum-2} we discuss a variational calculation that is related to the system of non-linear equations.

Two particular facts are important for any proposed factorization. Firstly, even if one can find factors ($\overline{\varphi}, \overline{\chi}$) both of which belong to the appropriate ${\cal{H}}^2$ Sobolev spaces, in the $r$ and $R$ variables respectively, it does not follow automatically that their product (\ref{hunxct}) belongs to ${\cal{H}}^2$ (in all variables $r,R$). Secondly it is possible that an electronic function $\phi(r,R)$ defined in the manner of Hunter, (\ref{hunewf}), does not belong to the Sobolev space ${\cal{H}}^2(r)$ and so cannot be interpreted as a bound-state electronic wavefunction. We give a model example in Appendix \ref{windmill}. Thus both approaches involve technical difficulties that must be overcome and both require a supplementary check that the product $\overline{\varphi}\overline{\chi}$ is actually an \textit{eigenfunction} of the Schr\"{o}dinger equation.

An alternative to Hunter's interpretation of the quotient $\phi(r,R)$, which we 
will also explore here, is to regard it as a \textit{molecular} wavefunction 
required to belong to the Sobolev space ${\cal{H}}^2(r,R)$. An example of 
such a factorization which largely avoids the troublesome technical details is 
given in Section ~\ref{agmon}. Although it lacks the direct physical 
interpretation of Hunter's approach it does incorporate precise information 
about the behaviour of eigenfunctions at infinity. In Section~\ref{norm-phy} 
we discuss the normalization of the factors proposed in 
\cite{AMG:10, AMG:12, LSC:13, LSC:14, H:75} in the light of the previous results, 
and correct the modified computation in\cite{LSC:14}; the relationship between 
the `electronic' and `molecular' wavefunction interpretations of $\phi$, 
equation(\ref{hunewf}), is discussed here. Finally we try to draw together 
our findings in Section~\ref{Disc}. An Appendix reviews  some key 
mathematical notions in an informal way.

\section{The Coulomb Hamiltonian}
\label{CouHam}
A molecule considered as a quantum mechanical collection of electrons and nuclei 
is customarily described by the usual Coulomb Hamiltonian $\mathsf{H}$ acting on 
an Euclidean configuration space defined by the particle coordinates with Schr\"{o}dinger equation
\begin{equation}
\mathsf{H}\Phi = E \Phi.
\label{Scheq}
\end{equation}
More explicitly, for a system of $N_e$ electrons and $N_n$ atomic nuclei we have
\begin{align}
\mathsf{H}&=\sum_{g=1}^{N_n}\frac{\mathsf{p}_g^2}{2m_g}+
\frac{e^2}{8\pi{\epsilon}_o}\sum_{g,h=1}^{N_n}\!\hbox{\raisebox{5pt}{${}^\prime$}}
\frac{Z_gZ_h}{r_{gh}^{nn}}\nonumber\\
&+\sum_{i=1}^{N_e}\left(\frac{\mathsf{p}_i^2}{2m}-\frac{e^2}{4\pi{\epsilon}_o}
\sum_{g=1}^{N_n}\frac{Z_g}{r_{ig}^{en}}\right)
+\frac{e^2}{8\pi
{\epsilon}_o}\sum_{i,j=1}^{N_e}\!\hbox{\raisebox{5pt}{${}^\prime$}}
\frac{1}{r_{ij}^{nn}}.
\label{hfull}
\end{align}
The configuration space is $\mathbb{R}^{3N_n+3N_e}$, where $\mathbb{R}$ denotes the set of real numbers. The interparticle distances are: $r_{gh}^{nn}=|{\bf x}_{g}^{n}- {\bf x}_{h}^{n}|, r_{ig}^{en}=|{\bf x}_{i}^{e}-{\bf x}_{g}^{n}|$, and $r_{ij}^{ee}=|{\bf x}_{i}^{e}-{\bf x}_{j}^{e}|$ in terms of particle coordinates \{${\bf x}_{i}^{e}, {\bf x}_{g}^{n}$\} in a laboratory frame. Here and elsewhere we use $(i,j)$ and $(g,h)$ as indices for electrons and nuclei respectively.

The primes on the summation symbols mean that terms with identical indices (`self-interactions') are to be omitted. We define the set $\Sigma_n$ of nuclear collisions as the set of those configurations such that $r_{gh}^{nn}=0$ for some $g, h$. Similarly, the set $\Sigma$ of all collisions is the set of those configurations such that $r_{gh}^{nn}=0$ for some $g, h$, or $r_{ih}^{en}=0$ for some $i,h$, or $r_{ij}^{ee}=0$ for some $i,j$. Collisions have important consequences for the analytical properties of the eigenfunctions which seem to have been first considered by 
Kato\cite{TK:57}.  In particular, it is expected that cusps appear in the wavefunction at collisions if the wavefunction does not vanish; it is also 
possible for an exact eigenfunction to have a node at the 
singularity\cite{PBB:66, MUSM:91}. Whether the inclusion of such cusps by means of trial wavefunctions involving $r_{ij}^{ee}$ in electronic structure calculations could  improve their accuracy has been quite widely studied; an example can be found  in \cite{PNFFG:01}. We shall consider the matter further from a mathematical standpoint\cite{SF:05} when examining the proposed factorisations.

For discussions of molecules where one wants to focus on the approximate separability of electronic and nuclear motions, it proves convenient to make a specific choice of the internal coordinates. The nuclear position variables \{${\bf R}^n$\} can be chosen as a set of $N_n-1$ translationally invariant variables, defined in terms of the original nuclear position coordinates, such that one of the new variables is the position coordinate for the nuclear centre-of-mass ${\bf X}$. The electronic coordinates \{${\bf r}^e$\} are a set of $N_e$ variables defined in terms of the original electronic coordinates by\cite{SW:05}
\begin{equation}
{\bf x}_{i}^{e}~=~{\bf r}_{i}^{e}~+~{\bf X}.
\label{ecoord}
\end{equation}

With this choice of coordinates the translationally invariant Coulomb Hamiltonian takes the form,
\begin{equation}
\mathsf{H}'\rightarrow {\mathsf H}^e({\mat r}^e) + {\mathsf H}^n
({\mat R}^n) + {\mathsf H}^{en}( {\mat R}^n, {\mat r}^e ).
\label{tich}
\end{equation}
The part of the Hamiltonian which can be associated with electronic
motion is
\begin{align}
\mathsf{H}^e({\mat r}^e) &= -\frac{\hbar^2}{2 m}\sum_{i=1}^{N_e}{\nabla}^2
({\mat r}^e_i) -\frac{\hbar^2}{2M_{N}}
\sum_{i,j=1}^{N_e}\vec{\nabla}
({\mat r}^e_i) \cdot \vec{\nabla}({\mat r}^e_j)\nonumber\\& +\frac{e^2}{8\pi
{\epsilon}_0}\sum_{i,j=1}^{N_e}\!\hbox{\raisebox{5pt}{${}^\prime$}}
\frac{1}{|{\mat r}^e_j-{\mat r}^e_i|}
\label{het}
\end{align}
where
\begin{equation}
M_{N}~=~\sum_{g=1}^{N_n}m_{g}.
\end{equation}

The part that can be associated with nuclear motion is
\begin{align}
\mathsf{H}^n({\mat R}^n) = &-\frac{\hbar^2}{2}\sum_{g, h=1}^{N_n-1}\frac{1}{\mu^n_
{gh}}\vec{\nabla}({\mat R}^n_g)\cdot\vec{\nabla}({\mat R}^n_h)\nonumber\\
&+ \frac{e^2}{8\pi{\epsilon}_0}\sum_{g,h=1}^{N_n}\hbox{\raisebox{5pt}
{${}^\prime$}} \frac{Z_g Z_h}{r_{gh}^{nn}({\mat R}^n)}
\label{hnt}
\end{align}
where $r_{gh}^{nn}({\mat R}^n)$ is the internuclear separation distance expressed in terms of the \{${\bf R}^n$\}
and the inverse mass matrix $1/ \mu^n_{gh}$ is in standard form\cite{SW:05}.

The electronic and nuclear motions are coupled only via a potential
term,
\begin{equation}
\mathsf{H}^{en}( {\mat R}^n, {\mat r}^e ) =
- \frac{e^2}{4\pi{\epsilon}_0}\sum_{g=1}^{N_n}\sum_{j=1}^{N_e}\frac{Z_g}
{r_{jg}^{en}( {\mat r}^e,{\mat R}^n)}
\label{hentc}
\end{equation}
where the electron-nucleus distance expression $|{\mat x}^{e}_{j} -{\mat x}^{n}_{g}|\equiv r_{jg}^{en}$ is again expressed in terms
of the internal coordinates. In the following it will be convenient to write \{${\bf r}^e$\} as $r$, and
\{${\bf R}^n$\} as $R$ for simplicity, and denote the gradient operator on nuclear coordinates as $\nabla^n$.
The first (sum) term in (\ref{hnt}) is the kinetic energy operator $\mathsf{T}_n$ in (\ref{Hint}); we write it
in this shorthand notation as 
\begin{equation}
\mathsf{T}_n = \frac{\hbar^2}{2\mu}\nabla^n\cdot\nabla^n.
\label{kinen}
\end{equation}
The operator $\mathsf{H}_{el}$ in (\ref{Hint}) is composed from the sum of the terms in (\ref{het}), (\ref{hentc}) and the last term in (\ref{hnt}).

The Schr\"{o}dinger equation for $\mathsf{H}'$ defined by (\ref{tich}) - (\ref{hentc}) is formally an elliptic partial differential equation (PDE) in the coordinates $(r,R)$,
\begin{equation}
\mathsf{H}'\Psi= E\Psi,
\label{ellip}
\end{equation}
on the reduced configuration space ${\cal X}=\mathbb{R}^{3N_n+3N_e-3}$. The occurrence of the Coulomb singularities in $\mathsf{H}'$ and the physical interpretation of $\Psi$ require that (\ref{ellip}) must be placed in a more general mathematical setting involving the notion of \textit{distributional derivatives} if it is to be given a precise meaning; we refer to Appendix \ref{distdiff} for details on distributional derivatives. One has to view (\ref{ellip}) in the following way.

Let us denote by ${\cal L}^2({\cal X})$ the set of square integrable functions on ${\cal X}$.
We define the Sobolev space ${\cal H}^2({\cal X})$ as the space of ${\cal L}^2({\cal X})$-functions such that their distributional derivatives up to second order all belong to ${\cal L}^2({\cal X})$.  For $\Psi\in {\cal H}^2({\cal X})$, each term in \eqref{ellip} makes sense as a ${\cal L}^2({\cal X})$-function and the equality takes place in this space ${\cal L}^2({\cal X})$. For instance, the term $\mathsf{T}_n\Psi$ is a ${\cal L}^2({\cal X})$-function that satisfies, for all smooth functions $h$ on ${\cal X}$ with bounded support,
\begin{equation}
\langle \mathsf{T}_n\Psi|h\rangle_{{\cal L}^2({\cal X})} = \langle \Psi|\mathsf{T}_n h \rangle_{{\cal L}^2({\cal X})},
\label{distn}
\end{equation}
where $\mathsf{T}_n h$ is now computed in the usual sense. One can see an eigenfunction $\Psi \in L^2({\cal X})$ as a \textit{distributional solution} to (\ref{ellip}). This means that for all smooth functions $h$ on ${\cal X}$ with bounded support, 
\begin{equation}
\langle \Psi|\mathsf{H}' h\rangle_{L^{2}({\cal X})} = E\langle \Psi|h \rangle_{L^{2}({\cal X})}.
\label{distn1}
\end{equation}

Essentially what is done here is the differentiations in $\mathsf{H}'$ are transferred to suitably smooth functions $h$, using integration by parts, as required. Note this point of view is already necessary in the simplest case: the Hydrogen atom. After removal of the centre-of-mass motion, the internal Coulomb Hamiltonian involves the electron-proton relative coordinate ${\bf r}$. The groundstate is given by $\Psi_0 = c\exp(-|{\bf r}|)$, in appropriate units. This function is continuous everywhere, and differentiable outside the collision at $0$. But it is not differentiable at 0 and so (\ref{ellip}) cannot be understood in the usual way. Now if the potential energy terms were smooth functions, for example Hooke's Law for coupled oscillators, the reformulation just described would yield (smooth) solutions everywhere that were solutions of the PDE (\ref{ellip}) in the usual sense. It is the occurrence of the singularities in the Coulomb potential that cause the main difficulties (to be discussed below) for the idea of an `exact factorization' of a molecular wavefunction. For future reference we denote ${\cal H}^2(\mathbb{R}^{3N_n-3})$ as the Sobolev space for wavefunctions depending on only the $N_{n}-1$ nuclear coordinates; it is contained in the corresponding space ${\cal L}^2(\mathbb{R}^{3N_n-3})$.

The best known regularity of an eigenfunction $\Psi$ of the Coulomb Hamiltonian $\mathsf{H}'$ is only that its first distributional derivatives are bounded\cite{TK:57, SF:05}; in particular, we do not know if one can differentiate $\Psi$ everywhere in the usual sense.  Fortunately, we have some further information on
 $\Psi$; we know by elliptic regularity\cite{LH:76} (see Appendix \ref{ellreg}), that the
following two statements are valid:
\begin{align}
\label{realanal}
\Psi~&\mbox{is a real analytic function outside the set}~\Sigma.\\
\label{zeroes}
\Psi~&\mbox{has at most isolated zeroes outside}~\Sigma.
\end{align}
If we replace the Coulomb interaction by some smooth potential, then elliptic regularity shows that $\Psi$ is smooth everywhere. This explains why the derivatives in (\ref{ellip}) can be computed in the usual sense in such a case.

These facts about a Coulomb eigenfunction already yield some useful information
about its putative factors. Thus, for example, if $\Psi$ can be written as 
$\overline{\chi}(R)\overline{\varphi}(r,R)$, then neither $\overline{\chi}$ nor 
$\|\overline{\varphi}\|_{r}^2:=\langle \overline{\varphi}, \overline{\varphi}\rangle_r$ can vanish outside $\Sigma _n$. Indeed, if $\overline{\chi}$ or $\|\overline{\varphi}\|_{r}^2$  vanishes at some $R_0$ then so does $\Psi$ on $\{(r, R); R=R_0\}$. If $R_0\not\in\Sigma _n$, then this set contains at least a segment outside $\Sigma$. This contradicts the fact, (\ref{zeroes}), that $\Psi$ has isolated zeroes 
outside $\Sigma$. On the other hand since we do not know if $\Psi$ has usual derivatives everywhere we cannot reasonably assume that the factors $\overline{\varphi}$ and 
$\overline{\chi}$ are everywhere regular. This information is directly relevant 
to our consideration of a system of equations that provide a \textit{formal} definition of factors $\overline{\varphi}$ and $\overline{\chi}$.

\section{Eigenvalue equation versus non-linear system}
\label{s:cederbaum-1}

In this Section we study the factorisation of eigenfunctions and the associated non-linear problem solved by the factors that were presented by Cederbaum\cite{LSC:13, LSC:14} and in the contributions of Gross \textit{et al.}\cite{AMG:10, AMG:12}. We also review Hunter's factorisation\cite{H:75}. To begin with we follow Cederbaum's arguments applied to an eigenfunction of $\mathsf{H}'$.

Firstly recall that Hunter started not from an assumption of `nuclear' and `electronic' factors but from an exact wavefunction for the molecular
system, which he then analysed in terms of conditional and marginal probability amplitudes to yield a factorization. Cederbaum's approach is rather different. In his equation (7a) it is \textit{assumed}\cite{LSC:13} that a product form $\overline{\varphi}(r,R)\overline{\chi}(R)$, where these functions are the putative solutions of a pair of coupled equations, can represent an exact wavefunction $\Psi$, rather than that the exact wavefunction can be written in product form; equation (7a) thus needs an existence proof.

Starting from a normalized solution $\Psi$ of $\mathsf{H}'\Psi =E\Psi$ and making the ansatz that $\Psi (r, R)$ factorises into $\overline{\varphi}(r,R)\overline{\chi}(R)$, one can try to follow Cederbaum's discussion (\S IIA\cite{LSC:13}), disregarding for the moment the question of normalization that will be studied later in Section~\ref{norm-phy}. Cederbaum's formal computation\cite{fn1} can be summarized as follows:
\begin{eqnarray}
\label{expansion}
0&=&(\mathsf{H}'-E)\overline{\varphi}\overline{\chi}\nonumber\\
0&=& \overline{\varphi}(\mathsf{T}_n\overline{\chi})-
\frac{\hbar^2}{\mu}\nabla^n\overline{\chi}
\cdot\nabla^n\overline{\varphi}+\overline{\chi}(\mathsf{H}'-E)\overline{\varphi}
\\
\label{factorisation}
0&=&\overline{\chi}\big(\overline{\chi}^{-1}(\mathsf{T}_n\overline{\chi})\overline{\varphi}-\frac{\hbar^2}{\mu}\overline{\chi}^{-1}\nabla^n\overline{\chi}
\cdot\nabla^n\overline{\varphi}\nonumber\\&~~~+&(\mathsf{H}'-E)\overline{\varphi}\big)
\\
0&=&\big(\mathsf{H}'- \frac{\hbar^2}{2\mu}\overline{\chi}^{-1}\nabla^n\overline{\chi}
\cdot\nabla^n-\overline{E}_{el}(R)\big)\overline{\varphi}
\label{dividing}
\end{eqnarray}
with
\begin{equation}
\overline{E}_{el}(R)\ =-\overline{\chi}^{-1}(\mathsf{T}_n\overline{\chi})+E.
\label{def-E-el}
\end{equation}

We can rewrite \eqref{dividing} as 
\begin{equation}
\overline{\mathsf{H}}_{el}\overline{\varphi}\ = \ \overline{E}_{el}(R)\overline\varphi
\label{eleqn-1}
\end{equation}
with
\begin{equation}
\overline{\mathsf{H}}_{el}\ = \ \mathsf{H}' - \frac{\hbar^2}{2\mu\overline{\chi}}{\bf \nabla}^n\overline{\chi}\cdot
{\bf \nabla}^n\, .\label{Hel-1}
\end{equation}
Multiplying \eqref{eleqn-1} by $\overline{\varphi}^\ast$ and integrating over the electronic variables, we get
\begin{equation}
{\overline{E}}_{el}(R) \|\overline\varphi\|_r^2\ = \ \langle{\overline\varphi}|\overline{\mathsf{H}}_{el}
\overline{\varphi}\rangle_{r}\, .
\label{elenergy-1}
\end{equation}
From \eqref{def-E-el} and \eqref{elenergy-1}, we derive
\begin{equation}
\mathsf{T}_n\overline{\chi}\ =\ \Bigl(E-\frac{\langle{\overline\varphi}|\overline{\mathsf{H}}_{el}
\overline{\varphi}\rangle_{r}}{\|\overline\varphi\|_r^2}\Bigr)\overline{\chi}\, .
\label{nucleqn-1}
\end{equation}
Now, if $(\overline{\varphi}, \overline{\chi})$ solves the coupled, non-linear equations \eqref{eleqn-1} and \eqref{nucleqn-1}, then, reversing the above computation, we get \eqref{expansion} and $\overline{\varphi}\overline{\chi}$ is a solution of
$(\mathsf{H}'-E)\overline\varphi\overline\chi=0$.

The present paper does not offer a detailed investigation of the non-linear equations (\ref{eleqn-1}) and (\ref{nucleqn-1}) but studies their relationship with the factorization. What is required is a framework in which the above computation can
actually be realized. A natural assumption to make would be that $\overline{\varphi}\in {\cal H}^2({\cal X})$ and $\overline{\chi}\in {\cal H}^2(\mathbb{R}^{3N_n-3})$, since one wants to interpret them as wavefunctions. Furthermore their product is
to be an eigenfunction, $\overline{\varphi}\overline{\chi}=\Psi$, and $\Psi\in{\cal H}^2({\cal X})$ is therefore essential. Since the latter property is \textit{not} guaranteed by $\overline{\varphi}\in {\cal H}^2({\cal X})$ and $\overline{\chi}\in {\cal H}^2(\mathbb{R}^{3N_n-3})$ (see Appendix \ref{prodH2}) one could try to 
study the non-linear equations in a subspace ${\cal H}_0$ of ${\cal H}^2({\cal X})$
\begin{equation}
{\cal H}_0\ =\ \{(\overline{\varphi},\overline{\chi})\in {\cal H}^2({\cal X})\times{\cal H}^2(\mathbb{R}^{3N_n-3}); \overline{\varphi}\overline{\chi}
\in {\cal H}^2({\cal X})\} .
\label{def-cal-H_0}
\end{equation}
On the other hand it is not at all obvious that all of the steps (\ref{expansion}) - (\ref{nucleqn-1}) in the computation are valid in the setting (\ref{def-cal-H_0}).

First of all, since $\overline{\chi}$ may have zeroes, the meaning of the division by $\overline{\chi}$ in the above formulae requires explanation. This is actually a delicate issue; to see this, let us take an example. Let $f$ be a smooth function on ${\cal X}$ with support in the region $\{|r|\leq 1; |R|\leq 1\}$ and such that $f=1$ for $(r, R)$ close to $0$. Let $g$ be a smooth function on $\mathbb{R}^{3N_n-3}$ with bounded support such that $g(R)=1$ for $|R|\leq 2$. Consider the smooth function $\overline{\chi}(R)=g(R)|R|^{2m}$, for some integer $m$. Away from $R=0$, $f/\overline{\chi}$ is a smooth function (given by $f(r, R)|R|^{-2m}$) but it is a quite nasty function near $R=0$ if $m$ is large enough\cite{fn3}. In particular, it does not belong to ${\cal{L}}^2({\cal X})$ and it is not clear how to interpret it as a distribution.

Thus we must consider how we might give a definite meaning to the computation \eqref{expansion}-\eqref{nucleqn-1}. In our first approach which is a local treatment, we avoid the set $\Sigma$, that is we restrict the values \{$r,R$\} to lie outwith $\Sigma$. In view of (\ref{realanal}) it would be natural to assume that the factors $\overline{\varphi}$ and $\overline{\chi}$ have at least conventional derivatives up to second order in this region. Since both $\overline{\chi}$ and $||\overline{\varphi}||_{r}$ are non-zero outside $\Sigma_n$ the calculations make sense pointwise at any point $(r,R)\not\in\Sigma$; all derivatives can be taken in the usual sense. \textit{If} we can find such factors\cite{fn4} $\overline{\varphi}$ and $\overline{\chi}$ in ${{\cal{H}}_0}$ with the further properties that neither $\overline{\chi}$ nor $||\overline{\varphi}||_{r}$ vanish outside $\Sigma_{n}$, and that equations \eqref{eleqn-1} and \eqref{nucleqn-1} are satisfied away from $\Sigma$, then following the computation backwards we obtain \eqref{expansion} outside $\Sigma$. Then $\mathsf{H}'\overline{\varphi}\overline{\chi} - E \overline{\varphi}\overline{\chi}$ belongs to ${\cal L}^2({\cal X})$, and we have shown that it is zero outside $\Sigma$, and so simply zero since $\Sigma$ has zero volume (is a set of zero measure). Thus $\overline{\varphi}\overline{\chi} \equiv \Psi$ is an eigenfunction of $\mathsf{H}'$.

Let us now try a global treatment of the computation. In \eqref{expansion}, we used the Leibniz rule for derivatives. We do not know if it is valid here since it is possible that one factor  contains a singular part (a non ${\cal L}^2({\cal X})$-part) which is compensated in another term. We would not be able to separate the terms but this is precisely what we must
do in \eqref{eleqn-1}-\eqref{Hel-1}. Assume that \eqref{expansion} is valid with each term in ${\cal L}^2({\cal X})$, possibly after restricting $(\overline{\varphi}, \overline{\chi})$ to a smaller subset ${\cal H}'$ of ${\cal H}_0$, (\ref{def-cal-H_0}). Now, we face the division problem in \eqref{factorisation}. Since $\overline{\varphi}\in {\cal H}^2({\cal X})$,
$(\mathsf{H}'-E)\overline{\varphi}$ is well-defined as a ${\cal L}^2({\cal X})$-function. So we should see an equality between ${\cal L}^2({\cal X})$-functions in \eqref{dividing}. Let us only
consider $\overline{\chi}^{-1}{\bf \nabla}^n\overline{\chi}\cdot {\bf \nabla}^n\overline{\varphi}$; if one views ${\bf \nabla}^n\overline{\varphi}$ as a distribution, there is the problem that the product of distributions is not generally defined, if one can identify $\overline{\chi}^{-1}{\bf \nabla}^n\overline{\chi}$ with a distribution. Instead, we might view ${\bf \nabla}^n\overline{\chi}$ and ${\bf \nabla}^n\overline{\varphi}$ as square integrable functions. Then ${\bf \nabla}^n\overline{\chi}\cdot {\bf \nabla}^n\overline{\varphi}$ is an integrable function in the $R$ variable. But the multiplication by $\overline{\chi}^{-1}$ may destroy this integrability property. Another try could be to see $\overline{\chi}^{-1}{\bf \nabla}^n\overline{\chi}\cdot {\bf \nabla}^n$ as a differential operator but, since $\overline{\chi}$ may vanish, it would be a singular one. Again, the result of its action on $\overline{\varphi}$ may be outside ${\cal L}^2({\cal X})$. Anyway, we see that one already has difficulties even to give a meaning to \eqref{factorisation}.

In Hunter's formulation\cite{H:75}, the first step of the computation (\ref{expansion}) - (\ref{nucleqn-1}) is performed for a special choice of $\overline{\chi}$, given by \eqref{hunf}.  The full computation is followed in the papers by Gross \textit{et al.}\cite{AMG:10, AMG:12, GG:14}. They do not require the nuclear wavefunction to be square integrable and so the removal of the centre-of-mass motion is not performed. However it is perfectly possible, and convenient, to discuss their method in the framework set out here. The nuclear function $\overline{\chi}$ is chosen as
\begin{equation}
\label{choice-chi}
\overline{\chi} (R)\ =\ e^{iS(R)}\Bigl(\int |\Psi (r, R)|^2\, \dd r\Bigr)^{1/2}\, , 
\end{equation}
where $S$ is an arbitrary real-valued function (cf (\ref{phase})) and $\Psi$ is  
a normalized solution of (\ref{ellip}). Then, $\overline{\varphi}$ is defined 
by $\Psi /\overline{\chi}$ and one derives equations for $\overline{\varphi}$ 
and $\overline{\chi}$ as above. Recall that they interpret $\overline{\varphi}$ 
as a $R$-dependent electronic wavefunction. The meaning of \eqref{expansion} already 
requires some information on the regularity of the factors. One can adapt 
the arguments in \cite{SF:05,fhhs1,fhhs2,J:10} to show that the function
\[R\ \mapsto \ \int |\Psi (r, R)|^2\, \dd r\]
is actually real analytic outside $\Sigma _n$. Since it does not vanish there, its square root is also real analytic and so is $\overline{\chi}$ in \eqref{choice-chi}, if $S$ is chosen real analytic. Away from $\Sigma$, the function $\overline{\varphi}$ defined by $\Psi /\overline{\chi}$ is therefore real analytic, by virtue of (\ref{realanal}).  Thus we can follow our `local' treatment performed above but, this time, we get a stronger result. The problem of factorisation of an eigenfunction of $\mathsf{H}'$ is {\it equivalent} to finding a solution $(\overline{\varphi}, \overline{\chi})$ of \eqref{eleqn-1} and \eqref{nucleqn-1} away from $\Sigma$ such that $\overline{\varphi}\overline{\chi}$ belongs to ${\cal H}^2({\cal X})$ and such that neither $\overline{\chi}$ nor $\|\overline{\varphi}\|_{r}$ vanish outside $\Sigma_n$.

The difficulties described above in the global approach also appear in the work 
of Gross \textit{et al.}\cite{AMG:10, AMG:12, GG:14} as we now show. Of course, we have $\Psi =\overline{\varphi}\overline{\chi}$ but, since the integral in \eqref{choice-chi} may vanish we have to be more precise in the definition of $\overline{\varphi}$. In view of (\ref{zeroes}) and repeating the argument at the end of Section~\ref{CouHam}, we see that the integral in \eqref{choice-chi} can vanish only in $\Sigma _n$; thus $\overline{\varphi}$ is well-defined outside $\Sigma _n$. This would be sufficient to define $\overline{\varphi}$ everywhere as $\Psi /\overline{\chi}$ if it were in ${\cal L}^2({\cal X})$. But the latter property is not certain since we do not know the behaviour of $\Psi/\overline{\chi}$ at the collisions, nor do we know if it is small enough at infinity for
$\overline{\varphi}$ to be square integrable. The same remarks apply to its
derivatives of course.

If one replaces the Coulomb interaction by a real analytic potential, then one can show by elliptic regularity that $\Psi$ is real analytic everywhere and the above discussion is valid with an empty set $\Sigma$. So the computation outside $\Sigma$ is actually the global one. Nevertheless, one would still have to characterize the behaviour at infinity of the product $\overline{\varphi}\overline{\chi}$, since it must belong to ${\cal H}^2({\cal X})$ if it is to be an eigenfunction of $\mathsf{H}'$.

\section{Variational method}
\label{s:cederbaum-2}

A classical way to find solutions of a partial differential equation is to 
introduce an appropriate functional such that its critical points are precisely 
the solutions of the given equation. Then one tries to find local extrema of 
the functional. Cederbaum proposed to follow this strategy 
and introduced a functional on functions $\overline{\varphi}(r,R)$ and 
$\overline{\chi}(R)$ and on two real parameters, having in mind that the 
product $\overline{\varphi}\overline{\chi}$ for a critical point should give 
a normalized eigenfunction of $\mathsf{H}$\cite{LSC:13}. Cederbaum's arguments 
do not prove the existence of such factorised eigenfunctions  since he does not 
prove the existence of critical points. There are important difficulties that 
make the search for critical points a delicate matter as we now describe. In the following we investigate the variational approach to the factorization of
eigenfunctions of the internal Hamiltonian $\mathsf{H}'$.

Let ${\cal E}$ be the set of the eigenvalues of $\mathsf{H}'$. Recall that ${\cal H}^2({\cal X})$ is the space of wavefunctions $\Psi$ (with centre-of-mass removed) of $N_n$ nuclei and $N_e$ electrons such that $\Psi$ and its distributional derivatives up to second order are all square integrable. We recall (cf (\ref{def-cal-H_0})) that ${\cal H}_0$ is the set of couples, $(\overline{\varphi}, \overline{\chi})$ where $\overline{\varphi}\in {\cal H}^2({\cal X})$, and $\overline{\chi}\in {\cal H}^2(\mathbb{R}^{3N_n-3})$ is a nuclear wavefunction, and such that the product
$\overline{\varphi}(r, R)\overline{\chi}(R)$ belongs to ${\cal H}^2({\cal X})$. Consider the functional $\tau : {\cal H}_0 \times \mathbb{R}^2 \rightarrow \mathbb{R}$
specified by\cite{LSC:13}
\begin{align}
\big((\overline{\varphi},\overline{\chi});(\lambda,\mu)\big)\ \mapsto & \langle \overline{\varphi}
\overline{\chi}|\mathsf{H}'(
\overline{\varphi}\overline{\chi})\rangle + \lambda\left(1-||\overline{\varphi}\overline{\chi}||^2\right)\nonumber\\
+&\mu\left(1-||\overline{\chi}||^{2}_{R}\right).
\label{def-tau}
\end{align}
Let ${\cal H}_1$ be a subset of ${\cal H}_0$ such that $\tau$ is differentiable on ${\cal H}_1\times \mathbb{R}^2$, and $(\overline{\varphi},\overline{\chi})\in {\cal H}_1$, $(\lambda,\mu)\in \mathbb{R}^2$. Then the following computations are valid
\begin{equation}\label{critical-point-1}
\frac{d\tau}{d\overline{\varphi}}\ =\ 2\overline{\chi}^\ast(\mathsf{H}'-\lambda )\overline{\varphi}\overline{\chi}
\, ,\ \frac{d\tau}{d\overline{\chi}}\ =\ 2\overline{\chi}\bigl(\langle\overline{\varphi}|(\mathsf{H}'-\lambda )
\overline{\varphi}\overline{\chi}\rangle_{r}-\mu\bigr)
\, .
\end{equation}
$((\overline{\varphi},\overline{\chi}); (\lambda,\mu))$ is a critical point precisely when both terms in (\ref{critical-point-1}) are zero and 
\begin{equation}\label{critical-point-2}
||\overline{\varphi}\overline{\chi}||\ =\ 1\, ,\ ||\overline{\chi}||_{R}\ =\ 1\, .
\end{equation}
In such a case,
\[\mu |\overline{\chi}|^2=\overline{\chi}^\ast\langle\overline{\varphi}|(\mathsf{H}'-\lambda )
\overline{\varphi}\overline{\chi}\rangle_{r}=\langle\overline{\varphi}|\overline{\chi}^\ast (\mathsf{H}'-\lambda
)\overline{\varphi}\overline{\chi}\rangle_{r}=0.\]
Since $\overline{\chi}$ is not identically zero, $\mu =0$.

Let ${\cal R}_{\overline{\chi}}=\{(r, R); \overline{\chi}(R)\neq 0\}$. Denote by $Z_{\overline{\chi}}$ the complement, that is the set of zeroes of $\overline{\chi}$. By the second equation in
\eqref{critical-point-2}, $Z_{\overline{\chi}}$ cannot be the whole space ${\cal X}$. Inside $Z_{\overline{\chi}}$, the product $\overline{\varphi}\overline{\chi}$ is zero so that
$(\mathsf{H}'-\lambda )\overline{\varphi}~\overline{\chi}=0$ there.
On the region ${\cal R}_{\overline{\chi}}$, $(\mathsf{H}'-\lambda )\overline{\varphi}\overline{\chi}=0$, since we consider a critical point. If we assume that the boundary of $Z_{\overline{\chi}}$ (or ${\cal R}_{\overline{\chi}}$) has zero volume we can show that $\overline{\varphi}\overline{\chi}$ is an eigenfunction of $\mathsf{H}'$ as follows. We know that it belongs to ${\cal H}^2({\cal X})$. Thus $(\mathsf{H}'-\lambda )\overline{\varphi}\overline{\chi}$ is well-defined and belongs to ${\cal L}^2({\cal X})$. The latter is zero on ${\cal R}_{\overline{\chi}}$ and on $Z_{\overline{\chi}}$. Given that the boundary of $Z_{\overline{\chi}}$ has a zero volume, $(\mathsf{H}'-\lambda )\overline{\varphi}\overline{\chi}=0$ holds true in ${\cal L}^2({\cal X})$ and, thanks to the second equation in \eqref{critical-point-2}, $\overline{\varphi}\overline{\chi}$ is an eigenfunction of
 $\mathsf{H}'$ and $\lambda\in {\cal E}$. Using again elliptic regularity, $\overline{\varphi}\overline{\chi}$ must be real analytic away from $\Sigma$. As already pointed out, this implies that $Z_{\overline{\chi}}\subset \Sigma_n$, which has a zero volume. We thus have shown that, if we have a critical point $((\overline{\varphi},\overline{\chi}); (\lambda, 0))$
of $\tau$ such that the boundary of $Z_{\overline{\chi}}$ has zero volume, then $\overline{\varphi}\overline{\chi}$ is an eigenfunction of $\mathsf{H}'$ with energy $\lambda$.

This variational method is related to the system \eqref{eleqn-1} and \eqref{nucleqn-1}. Let
$((\overline{\varphi},\overline{\chi}); (\lambda, 0))$ be a critical point of $\tau$. We assume further that $Z_{\overline{\chi}}\subset \Sigma_n$ and that $\overline{\chi}$ is smooth outside $\Sigma_n$. We know that $(\mathsf{H}'-\lambda )\overline{\varphi}\overline{\chi}=0$ and rewrite this away from $\Sigma$ as
\begin{equation}\label{=0}
0\ =\ \overline{\varphi}(\mathsf{T}_n\overline{\chi})+\overline{\chi}\overline{\mathsf{H}}_{el}\overline{\varphi}-\lambda
\overline{\varphi}~\overline{\chi}.
\end{equation}
Defining $\mathsf{E}_{el}$ by $\lambda -(\overline{\chi})^{-1}(\mathsf{T}_n\overline{\chi})$ away from $\Sigma_n$, we get \eqref{eleqn-1} outside $\Sigma$. Multiplying \eqref{=0} by $\overline{\varphi}^\ast$, integrating over the electronic variables, and dividing by $\overline{\chi}$, we recover \eqref{nucleqn-1} away from $\Sigma$. Using again the definition of $\mathsf{E}_{el}$, we arrive at \eqref{nucleqn-1}, outside $\Sigma$.

An obvious difficulty for the present variational method is the determination of the space ${\cal H}_1$, where the critical points of $\tau$ live. As discussed in
Appendix \ref{includ}, we think that ${\cal H}_1$ is strictly included in the space ${\cal H}_0$, which is already not easy to describe. Another difficulty is related to our assumption that the boundary of the set $Z_{\overline{\chi}}$ has zero volume. Without this assumption, we do not see how to justify that, for a critical point, the corresponding product $\overline{\varphi}\overline{\chi}$ is an eigenfunction of $\mathsf{H}'$. We can only hope that an appropriate study of the critical points or the solutions of the system \eqref{eleqn-1} and \eqref{nucleqn-1} justifies this assumption.

\section{A special factorization}
\label{agmon}

In the paper by Gross \textit{et al.}\cite{AMG:10} the factorisation (\ref{choice-chi}) is chosen as the starting point; we saw in Section~\ref{s:cederbaum-1} that it could produce singularities similar to those in Cederbaum's approach. Here we present another factorization based on an important result by Agmon\cite{SA:82, HS:00}, that does not have this drawback.

Let $\Psi$ be an eigenfunction of $\mathsf{H}'$ with energy $E$ which is isolated in the spectrum (this energy condition is satisfied for relevant situations in Chemistry). Then, one can deduce from Theorem 4.13 in Agmon\cite{SA:82} that there exists $c>0$ such that the function $\exp (c|(r, R)|)\Psi (r, R)$ belongs to ${\cal L}^2({\cal X})$.
Here $|(r, R)|$ denotes the norm of the vector $(r, R)$, that is, $(|r|^2+|R|^2)^{1/2}$. Setting
$\langle R\rangle=(1+|R|^2)^{1/2}$, we choose the nuclear factor in the form
\begin{equation}
\chi (R)=a\exp (-c'\langle R\rangle) ~\mbox{with}~ a>0 ~\mbox{and}~ c\geq c'\sqrt{2}>0. 
\label{agmonchi}
\end{equation}
Then $\chi\in {\cal H}^2(\mathbb{R} ^{3(N_n-1)})$, $\chi$ is smooth, real valued and positive everywhere. Choosing $a$ appropriately, we can ensure that $\|\chi\|_R=1$. Now we define $\varphi=\Psi /\chi$. Since we have pointwise
$\langle R\rangle\leq\sqrt{2}|R|\leq \sqrt{2}|(r, R)|$ then 
\[\exp (c'\langle R\rangle)|\Psi (r, R)|\ \leq \ \exp (c|(r, R)|)|\Psi (r, R)|\, \]
and $\varphi$ belongs to ${\cal L}^2({\cal X})$. If $\Omega$ is a bounded subset of
${\cal X}$, then $\varphi$ actually belongs to ${\cal H}^2(\Omega)$. But we do not
know if $\varphi\in {\cal H}^2({\cal X})$. In other words, the (distributional) derivatives of $\varphi$ up to second order are locally square integrable but we do not have enough control of their behaviour at infinity to ensure that they are globally square integrable.

These properties are however sufficient to allow us to repeat the arguments of Section~\ref{s:cederbaum-1} on any bounded subset $\Omega$ of ${\cal X}$, starting from $\Psi =\varphi\chi$, for then all terms are in ${\cal L}^2(\Omega)$. The nuclear wavefunction $\chi$ is smooth, and so the explicit function $\overline{E}_{el}(R)$ given by \eqref{def-E-el} is
also smooth, and bounded, and  \eqref{eleqn-1} even makes sense globally in ${\cal X}$, in the sense of distributions, and takes the form

\begin{align}
\Bigl(\mathsf{H}' + \frac{\hbar^2c'R}{2\mu\langle R\rangle}\cdot
{\bf \nabla}^n\Bigr)\varphi\ =&\ \Bigl(E\, +\, \frac{\hbar^2c'}{2\mu\langle R\rangle^3}\bigl(|R|^2
(c'\langle R\rangle +1)\nonumber\\ -& 3N_n\langle R\rangle^2\bigr)\Bigr)\varphi\, .
\label{eleqn-2}
\end{align}

Note also that, this time, the multiplication by $\chi (R)$ preserves the space ${\cal H}^2({\cal X})$. So, if we can find a nonzero solution $\varphi$ of \eqref{eleqn-2}, such that $\varphi\in {\cal H}^2({\cal X})$, we can reverse the computation in Section~\ref{s:cederbaum-1} to show that $\varphi\chi$ is an eigenfunction of $\mathsf{H}'$ with energy $E$.

We can also modify the variational method presented in Section~\ref{s:cederbaum-2} in the following way. Setting $\chi (R)$ as in (\ref{agmonchi}) above, we consider the functional $\tau '$ defined on ${\cal H}^2({\cal X}) \times \mathbb{R}$ by
\begin{equation}
(\varphi , \lambda )\mapsto \langle \varphi\chi|\mathsf{H}'(\varphi\chi)\rangle + \lambda\bigl(1-||\varphi\chi||^2\bigr)\, .
\label{def-tau'}
\end{equation}
In contrast to the functional $\tau$, $\tau'$ can be differentiated everywhere and 
\begin{equation}\label{critical-point-tau'}
\frac{d\tau'}{d\varphi}\ =\ 2\chi ^\ast(\mathsf{H}'-\lambda )\varphi\chi
\ ,\ \frac{d\tau'}{d\lambda}\ =\ 1-||\varphi\chi||^2\, .
\end{equation}
Now, since $\chi$ does not vanish, at a critical point $(\varphi , \lambda )$, the product $(\mathsf{H}'-\lambda )\varphi\chi = 0$. Since $(\mathsf{H}'-\lambda )\varphi\chi \in {\cal L}^2({\cal X})$ this shows that $\varphi\chi$ is a normalized eigenfunction of $\mathsf{H}'$ with energy $\lambda$, and that $\lambda\in {\cal E}$. As above, we can redo the
computation of Section~\ref{s:cederbaum-1} to get \eqref{eleqn-2}.  We can also look for a local extremum of $\tau '$ at $(\varphi , \lambda )$, since it must be a critical point. Thanks to the fact that $\chi$ is smooth and non-vanishing, we avoid the difficulties  encountered in Sections~\ref{s:cederbaum-1} and~\ref{s:cederbaum-2}. Indeed, we may forget about the collision set $\Sigma$ (although it reappears in the regularity properties of the
solution of \eqref{eleqn-2}), and the computations are (almost) elementary. In the variational approach based on $\tau '$, (\ref{def-tau'}), we avoid completely the obscure spaces ${\cal H}_0$ and ${\cal H}_1$ and the troublesome assumption about the boundary of $Z_{\overline{\chi}}$.

Note that, in the above description, we have a relatively free parameter, namely $c'$, that occurs in the definition of $\chi$ in (\ref{agmonchi}). We could try to let $\tau'$ depend on $c'$ as well in the variational approach. We do not know if this option facilitates the search for critical points.

\section{Normalization}
\label{norm-phy}

In this section, we discuss the normalization used in Gross \textit{et al.}\cite{AMG:10} and in Cederbaum\cite{LSC:13}. In particular, we rectify a modification proposed in Cederbaum's Erratum\cite{LSC:14}. In the framework of Section~\ref{s:cederbaum-1}, let us assume that $\Psi (r, R)=\overline{\varphi}(r,R)\overline{\chi}(R)$. Cederbaum's normalization requires, for all $R$, that
\begin{equation}\label{el-normalization}
\|\overline{\varphi}\|_{r}^2\ =\ \int |\overline{\varphi}(r, R)|^2\, \dd r\ =\ 1\, .
\end{equation}
This implies that, for all $R$, 
\[\int |\Psi (r, R)|^2\, \dd r\ =\ |\overline{\chi}(R)|^2\, .\]
Thus \eqref{choice-chi} is satisfied and the factorisation is the same as in Gross \textit{et al.}\cite{AMG:10}. Conversely, if we start with the latter, we have already pointed out that $\overline{\varphi}$ may be very irregular near the nuclear collisions. Fortunately, as seen at the end of Section~\ref{s:cederbaum-1}, $\overline{\chi}$ can only vanish at the nuclear collisions, that is, in a small region (a set of measure zero). Thus $\overline{\varphi}$ can be defined and we deduce that \eqref{el-normalization} holds true outside the nuclear collisions $\Sigma_n$. Note further that $\overline{\varphi}\not\in {\cal L}^2({\cal X})$ so it cannot be interpreted as a molecular wavefunction; instead it should be viewed as a $R$-dependent electronic wavefunction as in the BO approximation. The normalization (\ref{el-normalization}) does not essentially change our discussion on global computations in Section~\ref{s:cederbaum-1}. Now assume, as in our discussion outside the collisions set $\Sigma$, that we have found a solution $(\overline{\varphi}, \overline{\chi})$ of \eqref{eleqn-1} and \eqref{nucleqn-1} away from $\Sigma$ such that $\overline{\varphi}\overline{\chi}$ belongs to ${\cal H}^2({\cal X})$ and such that neither $\overline{\chi}$ nor $\|\overline{\varphi}\|_{r}$ vanishes outside $\Sigma_n$. Then $\Psi =\overline{\varphi}\overline{\chi}$ but \eqref{el-normalization} might be false. The factorisation
\begin{equation}
\Psi (r, R)\ =\ \frac{\overline{\varphi}(r, R)}{\|\overline{\varphi}\|_{r}(R)}\, \cdot\,
\|\overline{\varphi}\|_{r}(R)\overline{\chi}(R)
\label{new-factorisation}
\end{equation}
does satisfy the normalization condition but now the function $\overline{\varphi}(r, R)/\|\overline{\varphi}\|_{r}(R)$ might be very irregular near $\Sigma$. Thus, it is not clear that the normalization \eqref{el-normalization} can be satisfied.

Assuming that $(\overline{\varphi}, \overline{\chi})$, just as above, solves the
system \eqref{eleqn-1} and \eqref{nucleqn-1} outside $\Sigma$, we can derive the non-linear system that the factors in \eqref{new-factorisation} should satisfy, correcting in this way the corresponding computation in \cite{LSC:14}. We define $\varphi =\overline{\varphi}/\|\overline{\varphi}\|_{r}$ and $\chi =\overline{\chi}\|\overline{\varphi}\|_{r}$
(instead of
$\varphi =\overline{\varphi}\|\overline{\varphi}\|_{r}$ and $\chi =\overline{\chi}/\|\overline{\varphi}\|_{r}$ in \cite{LSC:14}). Away from $\Sigma$, $\|\varphi\|_{r}=1$ and we know that
$(\mathsf{H}'-E)\overline{\varphi}\overline{\chi}=0$, thus $(\mathsf{H}'-E)\varphi\chi = 0$, since $\varphi \chi =\overline{\varphi}\overline{\chi}$. By the computation \eqref{expansion} -  \eqref{nucleqn-1}, we arrive, still outside
$\Sigma$, at
\begin{equation}\label{eleqn-3}
\mathsf{K}_{el}\varphi\ = \ E_{el}(R)\varphi
\end{equation}
where
\[E_{el}(R)\ =\ -\chi^{-1}(\mathsf{T}_n\chi )+E\ \mbox{and}\ \mathsf{K}_{el}\ = \ \mathsf{H}' -
\frac{\hbar^2}{2\mu\chi}{\bf \nabla}^n\chi\cdot
{\bf \nabla}^n\, , \]
and
\begin{equation}\label{nucleqn-3}
\mathsf{T}_n\chi\ =\ \bigl(E-\langle\varphi |\mathsf{K}_{el}
\varphi\rangle_{r}\bigr)\chi\, .
\end{equation}

\section{Discussion}
\label{Disc}

In the previous sections, we have reviewed two schemes for the factorization of eigenfunctions of the molecular Hamiltonian, that were described in the papers\cite{AMG:10, AMG:12, GG:14, LSC:13, LSC:14, H:75}. We have seen that many ambiguities appear not only in the computations but also in the meaning of the results. Based on mathematical results (well-known in the mathematical physics community), we have extracted the main ideas of these methods and implemented them in a coherent framework, giving in this way a precise
meaning to the statements and partially justifying the computations. We also have provided in \S \ref{agmon} a new factorisation by using the methods in a different way. 

The main results can be summed up as follows, starting with the first method. If an eigenfunction $\Psi$ of (\ref{ellip}) can be factored into $\overline{\varphi}(r, R)\overline{\chi}(R)$ with sufficiently regular factors, then the latter must satisfy the non-linear system of equations \eqref{eleqn-1} and \eqref{nucleqn-1}, outside the set of collisions $\Sigma$. If one prescribes $\overline{\chi}$ to be a marginal of $\Psi$ (see \eqref{choice-chi}), one is led to the same conclusion, the factors being automatically regular enough. Conversely, if one has a solution $(\overline{\varphi}, \overline{\chi})$ of the non-linear system away from the collisions that satisfies some further conditions, the product $\overline{\varphi}(r, R)\overline{\chi}(R)$ is an eigenfunction. If one chooses at the outset $\overline{\chi}$ as in equation (\ref{agmonchi}) one essentially gets the previous results without caring about the collisions. 

The second method is of variational nature. Having in mind to factorize an eigenfunction $\Psi$ as $\overline{\varphi}(r,R)\overline{\chi}(R)$ one introduces a functional, acting on functions $\overline{\varphi}$ and $\overline{\chi}$, that is defined on an appropriate but quite complicated space. The functional is chosen such that its critical points produce an eigenfunction $\Psi$ in the desired product form provided they satisfy some regularity condition. Furthermore, one can relate this approach to the previous non-linear system outside the collisions. In order to progress one must better understand the functional $\tau$ and the complicated sets ${\cal H}_0$ and ${\cal H}_1$.  In Appendix \ref{includ} we show that ${\cal H}_1$ is not empty and is probably strictly included in ${\cal H}_0$. A demonstration that the functional $\tau$ actually has critical points is a delicate matter. Furthermore we need an unpleasant assumption on the zero set of $\overline{\chi}$ to construct an eigenfunction from a critical point. When the factor $\overline{\chi}$ is chosen as (\ref{agmonchi}) one can also follow a similar variational method on a quite natural, simple space. 

For both approaches the results are quite limited. In the factorisation with $\overline{\chi}(R)$ given by a marginal, the other factor $\overline{\varphi}$ is smooth away from the collisions but we cannot, as yet, exclude an irregular behaviour near the collisions, which would not be compatible with an interpretation of their product as a wavefunction. Starting from a solution of the non-linear system, we do not know if the conditions required to prove that the product is indeed an eigenfunction are satisfied. In the variational method, we do not know if the functional has critical points; we even have difficulties to describe the space where we have to look for them. If we have such a critical point, we still need to check further properties to ensure that the product of factors is indeed an eigenfunction.

The situation is a bit better when one requires $\overline{\chi}$ to be specified by (\ref{agmonchi}). In this case, the non-linear system reduces to a linear equation for the other factor $\overline{\varphi}$ (see \eqref{eleqn-2}) and we just have to find a solution of this equation in a natural space (i.e. ${\cal H}^2({\cal X})$). In this framework, the variational method is also easier to work with. We know that the functional has critical points but we do not know how to compute them.

In the previous sections, we have seen that the set of collisions $\Sigma$ plays an important r\^ole. This is due to the fact that the Coulomb interactions have singularities precisely on this set. These collisions are responsible for most of the difficulties we encountered. If we regularize each Coulomb singularity, that is, make the replacement of each $x_{kl}^{-1}$ in the operator $\mathsf{H}'$ by a real analytic function, then an eigenfunction $\Psi$ is everywhere analytic. If we assume that $\Psi (r,R)= \overline{\chi}(R)\overline{\varphi}(r, R)$ or if we write this factorisation with $\overline{\chi}$ defined by \eqref{choice-chi}, we can show as above that $\overline{\chi}$ never vanishes. All the previous difficulties related to the behaviour of $\overline{\varphi}$ near $\Sigma$ and those of $\overline{\chi}(R)$ near $\Sigma _n$ disappear after the regularization. Therefore the regularized model presented in Gross \textit{et al.}\cite{AMG:10} and the exactly solvable one chosen by Cederbaum\cite{LSC:13} are not capable of giving insight into the actual molecular case, because an essential ingredient is lacking from the proposed models.

Given a factorization of a particular eigenfunction $\Psi(r,R)_i = f(R)_{i}\phi(r,R)_i$, a `non-adiabatic energy surface' for the nuclei can be defined formally by integrating out the electronic variables in the expectation value of the internal Hamiltonian in the state $\phi_{i}$, (cf (\ref{elenergy-1}))
\be
\mathsf{U}(R)_{i}= \int\phi(r,R)^{*}_{i}\mathsf{H}'(r,R) \phi(r,R)_{i}\dd r
\label{hunpes}
\ee
where $\mathsf{H}'$ is the molecular Hamiltonian with the centre-of-mass contribution removed (see (\ref{Hint})). Hunter showed that one can derive a `reduced Schr\"{o}dinger equation' for the nuclear function $f(R)_{i}$ that partners $\phi(r,R)_{i}$ in which $\mathsf{U}(R)_{i}$ appears as a potential energy contribution.

However it is important to note that this is \textit{not} a refinement of the conventional Schr\"{o}dinger equation for nuclear motion on a PES, because here the energy $E$ is \textit{fixed} (it is the eigenvalue of the specified eigenfunction $\Psi_{i}$). One can require $f(R)_{i}$ to vanish at the collisions ($R=0$), but the behaviour at $\infty$ is not \textit{a priori} assured. In the conventional adiabatic BO treatment $E$ along with the nuclear wavefunctions are unknowns, and one finds in the well-known way that satisfying the boundary conditions at $R=0,R=\infty$ to assure square integrability is only possible for certain discrete values of $E$, the molecular vibration-rotation levels associated with the PES.

The pseudo-potential $\mathsf{U}$ defined in (\ref{hunpes}) as introduced by Hunter\cite{H:75}, and studied computationally by Czub and Wolniewicz\cite{CW:78}, is only defined in a purely formal sense since, as we have seen (Section~\ref{s:cederbaum-1}), the function $\phi$ may be so irregular that the application of $\mathsf{H}'$ to it could be ill-defined. It would be sufficient however that, for fixed $R$, $\phi$ lies in the Sobolev space ${\cal H}^2$ in the $r$ variables, but we do not see what would guarantee such a property \textit{a priori}. Our analysis shows that, outside the collisions, $\phi$ is analytic though that is not enough to control the behaviour at large $r$. However, making use again of arguments in \cite{SF:05,fhhs1,fhhs2, J:10}, one might hope to show that, for fixed $R$ away from the nuclear collisions, $\phi$ does have the regularity ${\cal H}^2$ in the $r$ variables, so giving a meaning to (\ref{hunpes}), and also that $\mathsf{U}$ is analytic in this region.

Hunter thought it unlikely that a nuclear function $f(R)_{i}$ would have zeroes in view of his interpretation of it as a marginal probability amplitude function for the nuclear coordinates, but originally based his claim on an analysis of the Schr\"{o}dinger equation for coupled harmonic oscillators\cite{GH:74}. That problem is no real guide to the properties of the solution of the Schr\"{o}dinger equation for the Coulomb Hamiltonian not least because the oscillator Hamiltonian is separable, has purely discrete spectrum and is well-behaved at collisions.

Nevertheless, in the chemical physics literature on factorization of molecular wavefunctions it has been argued in more general terms that a nuclear 
wavefunction defined as a marginal probability amplitude for an 
exact eigenfunction, $\Psi$, as in (\ref{hunf}) or 
(\ref{choice-chi}), is necessarily \textit{nodeless}. The argument rests on 
the statement that there exists a set of orthonormal functions \{$\sigma_n(r,R)$\}
that is `complete in the adiabatic electronic space'; for definiteness, assume the
\{$\sigma_n$\} are the eigenfunctions of the clamped-nuclei Hamiltonian,
\begin{displaymath}
\mathsf{H}' - \mathsf{T}_n \rightarrow \mathsf{H}_{cn}
\end{displaymath}
in which the nuclear positions \{$R$\} are treated as classical parameters,
\begin{displaymath}
\mathsf{H}_{cn}(\mathsf{r}:R)\sigma_n(r,R)= e_n(R)\sigma_n(r,R)
\end{displaymath}
Then one writes an exact eigenfunction (of $\mathsf{H}'$) as in the Born-Huang
theory\cite{SW:12}
\begin{equation}
\label{BHeq}
\Psi(r,R) = \sum_n \chi_n(R)\sigma_n(r,R)
\end{equation}
and the nodeless property of the \{$\chi_n$\} follows\cite{GG:14, GH:81, CW:78}.

By contrast we are unable to exclude the possibility of nodes in the $\chi(R)$
functions for $R$ values associated with the collisions. Why the difference ?
The essential point is that an expansion such as (\ref{BHeq}) relies on the set
of eigenfunctions \{$\sigma_n$\} providing a \textit{resolution of the identity}, 
and this is only valid in the case of a purely discrete spectrum; in other words,
\{$\sigma_n$\} must be true eigenfunctions associated with (discrete) eigenvalues.
When an operator also (or only) has a continuous portion of spectrum the matrix 
notion of diagonalization providing a complete set of states breaks down, and the 
resolution of the identity must instead be developed from the spectral theorem 
and the idea of spectral projection. One of us has explored this idea\cite{TJ:14} 
in detail in the context of the BO approximation; we refer to that discussion 
which shows that an exact representation of $\Psi$ is much more complicated than
(\ref{BHeq}), and does not lead us to such a definite conclusion about the 
nodal properties of $\chi$.
 
So far, nothing has been said about spin statistics. Consider a collision 
involving two identical nuclei (1,2). Under a permutation $\mathsf{P}_{12}$ 
an exact eigenfunction $\Psi$ will either be symmetric (boson statistics) or
antisymmetric (fermion statistics). In the later case the eigenfunction
vanishes at the collision $R_{12}=0$, and a $\chi$ factor calculated according
to (\ref{hunf}) will also vanish. Direct examination of the Schr\"{o}dinger
differential equation in the vicinity of $R_{12}=0$ shows that the spatial
part of the wavefunction may vanish in any case\cite{PBB:66, MUSM:91}.

Based on the present knowledge, we have the following impression of these methods. The system \eqref{eleqn-1} and \eqref{nucleqn-1} is non-linear and has \textit{a priori} singularities. This is already a difficult problem, but
here we have the unusual situation in the case of \eqref{eleqn-1} where the singularities of the equation depend on an unknown function. Concerning the variational method, the space on which we can apply it is difficult to describe.
This comes precisely from the fact that the functional contains products of the variables, which are \textit{a priori} less regular functions. When one requires that the factor $\overline{\chi}$ or $\chi$ is given by an appropriate exponential function, as in Section~\ref{agmon}, the situation is a bit better but we do not see a real improvement in \eqref{eleqn-2} compared to the original equation \eqref{ellip}. We do not see a natural physical interpretation for the factors in this setting and judge the factorisation artificial. For these reasons, we are not convinced of the efficiency of the methods to produce eigenfunctions. It would seem that the equation systems suggested in these proposals are so difficult to handle that a direct approach to constructing eigenfunctions treating the electrons and nuclei on the same footing might be no more challenging; after all, the equations for the `electronic' factor $\varphi$ (or $\overline{\varphi}$) still contain the full internal molecular Hamiltonian $\mathsf{H}'$ and all the electronic \textit{and} nuclear variables of the problem.

\newpage

\section{Appendix}\label{appendix}

In this Appendix we explain some notions and results used in the main text.

\subsection{Differentiation in the distributional sense and products} 
\label{distdiff}
Let $d$ be an integer and $f$ be a locally integrable function on $\mathbb{R}^d$.
This means that, for any bounded subset $\Omega$ of $\mathbb{R}^d$, $f$ is integrable on $\Omega$. We denote
by ${\cal D}(\mathbb{R}^d)$ the space of smooth (complex-valued) functions on $\mathbb{R}^d$ with bounded support. One
can identify the function $f$ with the distribution $T_f:{\cal D}(\mathbb{R}^d)\to \mathbb{C}$ (the set of complex numbers) defined by
\[T_f(g) = \int_{\mathbb{R}^d}f(x)g(x)\, \dd x \]
The distributional derivative of $f$ w.r.t. $x_1$ is the corresponding derivative of $T_f$, which is the new distribution $\partial_{x_1}T_f:{\cal D}(\mathbb{R}^d)\to \mathbb{C}$ defined by 
\[\partial_{x_1}T_f(g) = -\int_{\mathbb{R}^d}f(x)\partial_{x_1}g(x)\, \dd x .\]
Note that, for $g\in{\cal D}(\mathbb{R}^d)$, $\partial_{x_1}g\in{\cal D}(\mathbb{R}^d)$. If $h$ is a smooth function on $\mathbb{R}^d$ , then $hg\in{\cal D}(\mathbb{R}^d)$, if $g\in{\cal D}(\mathbb{R}^d)$. If $T$ is a distribution on $\mathbb{R}^d$, that is a continuous (in an appropriate sense) linear map from ${\cal D}(\mathbb{R}^d)$ to $\mathbb{C}$, one defines the product of $T$ by the smooth function $h$ as the new distribution given by $(hT)(g)\ =\ T(hg)$. In particular, $hT_f=T_{hf}$.

Let us take examples that are relevant for the main text. For simplicity, we assume $N_e=N_n=1$. Let $\chi(R)\in {\rm L^2}(\mathbb{R}^3)$ and $\varphi (r,R)\in {\rm L^2}(\mathbb{R}^6)$. The distributional derivative of $\chi$ w.r.t. $R_1$ is the linear map
\[\tilde\partial _{R_1}\chi : {\cal D}(\mathbb{R}^3)\ni g\, \mapsto \, -\int_{\mathbb{R}^3}\chi(R)\partial_{R_1}g(R)\, \dd R .\]
We may multiply it by a smooth function. But the product by $\varphi (r, \cdot)$ (with fixed $r$) is \textit{a priori} undefined since
$\varphi (r, \cdot)g\in {\cal D}(\mathbb{R}^3)$ could be false. Indeed $\varphi (r, \cdot)$ could belong to the set of ${\rm L^2}(\mathbb{R}^3)$-functions that are not smooth. So, we have difficulty defining $\varphi (r, \cdot)\tilde\partial _{R_1}\chi$. However, if $\tilde\partial _{R_1}\chi=
T_{f_1}$, for a function $f_1\in {\rm L^2}(\mathbb{R}^3)$, we may define the product $\varphi (r, \cdot)\tilde\partial _{R_1}\chi$ as the usual
product $\varphi (r, \cdot)f_1$, which is in ${\rm L^1}(\mathbb{R}^3)$. We could also differentiate this product in the distributional sense since one can identify a ${\rm L^1}(\mathbb{R}^3)$-function with a distribution. If we would do that, the Leibniz rule
\[\tilde\partial _{R_1}\bigl(\varphi (r, \cdot)f_1\bigr)\ =\ (\tilde\partial _{R_1}\varphi (r, \cdot))f_1+\varphi (r, \cdot)
(\tilde\partial _{R_1}f_1)\]
might be false. Both products on the r.h.s might be undefined. This simple situation illustrates some difficulties mentioned in Section~\ref{s:cederbaum-1}.

\subsection{Products of ${\cal H}^2$-functions} 
\label{prodH2}
Here we construct functions $\varphi\in {\cal H}^2(\mathbb{R}^{3(N_e+N_{n-1})})$ and
$\chi\in {\cal H}^2(\mathbb{R}^{3N_{n-1}})$ such that $\varphi\chi\not\in {\cal H}^2(\mathbb{R}^{3(N_e+N_{n-1})})$. Let $f$ and $g$ be smooth functions with bounded support in $\mathbb{R}^{3(N_e+N_{n-1})}$ and $\mathbb{R}^{3N_{n-1}}$, respectively, such that both are equal to $1$ near $0$. Let $\alpha$ and $\beta$ be real numbers. We set 
\[\varphi (r, R)\ =\ f(r, R)\cdot \bigl(|r|^2+|R|^2\bigr)^{\alpha /2}\ \mbox{and}\ \chi (R)\ =\ g(R)\cdot |R|^\beta\, .\]
Then $\varphi\in {\cal H}^2(\mathbb{R}^{3(N_e+N_{n-1})})$ and $\chi\in {\cal H}^2(\mathbb{R}^{3N_{n-1}})$ if
\[\alpha \, >\, 2-\frac{3}{2}(N_e+N_{n-1})\ \mbox{and}\ \beta\, >\, 2-\frac{3}{2}N_{n-1}\, .\]
Provided that $N_e\geq 3$, one can choose $\alpha$ and $\beta$ satisfying the above conditions and also $\alpha +\beta\leq -3N_{n-1}$. The latter implies that $\varphi\chi\not\in {\cal L}^2(\mathbb{R}^{3(N_e+N_{n-1})})$ and thus $\varphi\chi\not\in {\cal H}^2(\mathbb{R}^{3(N_e+N_{n-1})})$.

\subsection{${\cal H}_1\neq {\cal H}_0$ ?}
\label{includ}

To begin with, if $\varphi$ and $\chi$ are smooth functions with bounded support in
$\mathbb{R}^{3(N_e+N_n)}$ and $\mathbb{R}^{3N_n}$, respectively, then $(\varphi , \chi )\in {\cal H}_1$. Thus ${\cal H}_1$ is not empty.
By definition, ${\cal H}_1\subset {\cal H}_0$ but even so we expect that one can find
$(\varphi , \chi )\in {\cal H}_0$ at which the functional $\tau$ is not continuous (and thus not differentiable). To motivate this guess, we shall prove it for a simpler functional related to $\tau$ (Section~\ref{s:cederbaum-2}). Consider the map $\tau _0:{\cal H}_0\rightarrow \mathbb{R}$ given by 
\[(\overline{\varphi},\overline{\chi})\ \mapsto \ ||\overline{\varphi}~\overline{\chi}||^2\, ,\]
which is a part of the functional $\tau$. Take $N_n\geq 3$. Let $\beta =2-3N_n/2+1/8>2-3N_n/2$. Let $\delta =3N_n/2+\beta +1/8=2+1/4$. In particular,
$2+\delta <3N_n/2$. Let $f$ be a non-zero, smooth function on $\mathbb{R}^{3N_e}$ with bounded support. Let $g$ and $h$ be two smooth functions on $\mathbb{R}^{3N_n}$ with bounded disjoint supports such that $g=1$ near zero. Since the supports are disjoint, $gh$ is identically zero. Let
\[\varphi (r, R)\ =\ f(r)\cdot h(R)\ \mbox{and}\ \chi (R)\ =\ g(R)\cdot |R|^\beta\, .\]
Thanks to $\beta >2-3N_n/2$, $(\varphi , \chi)$ belongs to ${\cal H}_0$; obviously $\varphi\chi =0$ identically.
For all integer $j$, let $g_j(R)=j^\delta g(jR)$. Notice that, for $j$ large enough, the support of $g_j$ is included in the region about $0$ where $g=1$. Let 
\[\varphi _j(r, R)\ =\ f(r)\cdot g_j(R)\, .\]
Since $2+\delta <3N_n/2$, we see that $g_j$ goes to $0$ in ${\cal H}^2(\mathbb{R}^{3N_n})$ and thus $\varphi _j$ goes to $0$ in ${\cal H}^2(\mathbb{R}^{3(N_n+N_e)})$ as $j$ goes to infinity. 

If $\tau _0$ were continuous at $(\varphi , \chi)$ then the difference $\tau _0(\varphi +\varphi _j, \chi)-\tau _0(\varphi , \chi)$
should go to $0$, as $j$ goes to infinity. This difference is given by
\[\tau _0(\varphi +\varphi _j, \chi)-\tau _0(\varphi , \chi)\ =\ \|\varphi _j\chi\|^2\, ,\]
since $\varphi\chi =0$. Now, for $j$ large enough,
\begin{align*}
\|\varphi _j\chi\|^2\, =&\ \|f\|_r^2\cdot \int_{\mathbb{R}^{3N_n}}|g_j(R)|^2\cdot |R|^{2\beta}\, \dd R\\
=&\ j^{2\delta -3N_n-2\beta}\|f\|_r^2\cdot \int_{\mathbb{R}^{3N_n}}|g(s)|^2\cdot |s|^{2\beta}\, \dd s
\end{align*}
and the r.h.s. blows up as $j$ goes to infinity, since $2\delta -3N_n-2\beta =1/4>0$.
This yields a contradiction, showing that $\tau _0$ is not continuous at $(\varphi , \chi)$. 

\subsection{$\phi$ not necessarily in ${\cal{H}}^2$}
\label{windmill}

The point we want to demonstrate here is simply that 
one must make a check to ensure that a factor 
$\phi(r;R)$, defined by (\ref{hunewf}), belongs to the Sobolev 
space ${\cal{H}}^2(r)$. This is an essential property 
if $\phi$ is to be interpreted as a wavefunction for the electrons for fixed $R$. 
Here we propose two model examples that show that $\Psi/f$ is not automatically in
${\cal{H}}^2$ (in the $r$ variables) when $\Psi$ belongs to ${\cal{H}}^2$ (in 
all variables, $r,R$) and $f$ is defined as a marginal as in (\ref{hunf}). In 
the first model $\Psi/f$ does not belong to ${\cal{H}}^2$ because of irregular 
behaviour near $R=0$; in the second the irregular behaviour is located at $\infty$ 
(in the $R$ variable).

We choose  a smooth, nonnegative function $\tau$ of one 
real variable such that
\begin{equation*}
\begin{split}
&\tau(t)=\left\{\begin{array}{ll}1~~~\mbox{if}~|t|~\leq~\frac{1}{2}\\
0~~~\mbox{if}~|t|~\geq~1\end{array}\right. \nonumber
\end{split}
\end{equation*}
For simplicity we choose three dimensional variables ${\bf r}$ and ${\bf R}$ with 
$r=|{\bf r}|,R=|{\bf R}|$. The configuration space $\Omega=\mathbb{R}^6$.
Let 
\begin{equation}
\Psi({\bf r},{\bf R}) = R^n~\tau(R)~\tau(R^{2m}(r~-~R))\nonumber
\end{equation}
for some integers $n > 0, m < 0$. One can specify $n,m$ such that this 
`wavefunction' $\Psi$ belongs to the Sobolev space ${\cal{H}}^2(\Omega)$ (essentially $|n|$ has to be large compared to $|m|$). 

 Following the prescription of Hunter\cite{H:75} we then have, after integration over the angles and some simplification, for some $c>0$, 
\begin{align}
f(R)&= cR^{n+1-m} \tau(R) \cdot\left(\int_{\mathbb{R}}\tau(t)^2\left(1+tR^{-2m-1}\right)^2\dd t\right)^{\tfrac{1}{2}}\,\nonumber .
\end{align}
Note that the vanishing of $f$ at $R=0$ is stronger than that of $\Psi$. Now we define 
\begin{equation}
\begin{split}
&\phi(r;R)=\left\{\begin{array}{ll} 0\ \mbox{if}\ \tau (R)=0\ \mbox{else}\\
\frac{\Psi(r,R)}{f(R)}\, \end{array}\right. \nonumber.
\end{split}
\end{equation}
In particular, we do not define $\phi$ on the zero volume region $\{R=0\}$. 
Using the same changes of variables as above, we get for some $c'>0$, 
in the region where $\tau (R)\neq 0$, 
\begin{align}
\phi(r;R)= c'R^{m-1}& \tau\big(R^{2m}(r-R)\big)\cdot\nonumber\\&\left(\int_{\mathbb{R}}\tau(t)^2\left(1+tR^{-2m-1}\right)^2\dd t\right)^{-\tfrac{1}{2}}\nonumber
\end{align}
Explicit calculation shows that the `electronic' function $\phi$ is 
square integrable, but that its first derivative (in the $r$-variable) is not for 
$|m|$ large enough, so that it no longer belongs to ${\cal{H}}^2$. Nevertheless 
there are $n$ values such that $\Psi$ is in ${\cal{H}}^2(\Omega )$ ( $n > -7m/2-2$).
The reason for this behaviour comes from the fact that $f$ vanishes more 
strongly at $R=0$ compared to $\Psi$.

We can use this idea again to translate the irregular behaviour at $R=0$ to 
$R=\infty$ . Taking 
\begin{equation}
\Psi({\bf r},{\bf R}) = R^n~(1-\tau(R))~\tau(R^{2m}(r~-~R))\nonumber
\end{equation}
but now for large $-n$ and positive $m$, we can check that $\Psi$ belongs to 
the Sobolev space ${\cal{H}}^2(\Omega)$ (essentially $|n|$ has again to be 
large compared to $|m|$). Again we can adjust $m$ such that the 
${\bf r}$-gradient of $\phi$ is not square integrable.

\subsection{Local Sobolev space on $\mathbb{R}^{3(N_e+N_n)}$} 
\label{Sobspace}
In the text, we defined the Sobolev space ${\cal H}^2(\mathbb{R}^{3(N_e+N_n)})$. Similarly,
we can define ${\cal H}^2(\Omega)$, for any bounded open subset $\Omega$ of $\mathbb{R}^{3(N_e+N_n)}$. The corresponding local Sobolev space,
denoted by ${\cal H}^2_{{\rm loc}}(\mathbb{R}^{3(N_e+N_n)})$ is the space of those
functions $f$ that belong to ${\cal H}^2(\Omega)$, for all bounded open subsets $\Omega$ of $\mathbb{R}^{3(N_e+N_n)}$. A function
that has usual continuous derivatives up to second order always belongs to the local Sobolev space ${\cal H}^2_{{\rm loc}}(\mathbb{R}^{3(N_e+N_n)})$,
but the integral of its modulus square can be infinite (just think about the constant function equal to one). Therefore the
local and global ${\cal H}^2(\mathbb{R}^{3(N_e+N_n)})$-spaces are different, the latter being included in the former. 

\subsection{$\mathsf{H}$ has no eigenvalue } 
\label{noeig}
Let us first give a `physical proof'. Since the full molecular system is considered as being isolated in the universe, its mass centre is freely moving. Thus it cannot be in a bound state. 

Now we turn to a mathematical proof. We start with \eqref{Hint} (assuming for simplicity that $\hbar^2/2M_T=1$) and recall that the internal Hamiltonian $\mathsf{H}'$ is ${\bf R}$-independent.
Now we view this formula in the Fourier space of ${\bf R}\in \mathbb{R}^3$. Denoting by $\mathsf{H}_1$ this representation of $\mathsf{H}$, we get
$\mathsf{H}_1=\mathsf{H}'+\mathsf{M}_f$, where $\mathsf{M}_f$ is the multiplication operator by the function $f(\xi )=|\xi |^2$ ($\xi$ being
the Fourier variable associated to ${\bf R}$). This can be rewritten with the help of a direct integral (cf. \cite{RS4:78}, p. 279-287)
as
\[\mathsf{H}_1\ =\ \int_{\mathbb{R}^3}^\oplus\bigl(|\xi |^2\, +\, \mathsf{H}'\bigr)\, \dd\xi .\]
According to \cite{RS4:78}, p. 284, $E$ is an eigenvalue of $\mathsf{H}_1$ (or $\mathsf{H}$) if and only if
the $3$-dimensional volume of $C_E=\{\xi\in \mathbb{R}^3; E\, \mbox{is an eigenvalue of}\, |\xi |^2+\mathsf{H}'\}$ is positive.
Note that $E$ is an eigenvalue of $|\xi |^2+\mathsf{H}'$ if and only if $E-|\xi |^2$ is an eigenvalue of $\mathsf{H}'$.
By the Mourre theory (see \cite{ABG:96}), one can show that the set of eigenvalues of $\mathsf{H}'$ is at most countable. 
This implies that $C_E$ is the union of an at most countable set of spheres. Thus $C_E$ has zero volume and $\mathsf{H}$ has no eigenvalue.

\subsection{Elliptic regularity} 
\label{ellreg}
We propose here a short and intuitive introduction to elliptic regularity.
For more details, we refer to \cite{LH:76, RS2:75}.

We consider the differential equation $\nabla ^2u=f$ on $\mathbb{R} ^2$, where the function $f$ is given and $u$ is the unknown function. Our goal is, knowing the regularity properties of $f$, to obtain those of any solution $u$. In one dimension, the problem is easy, since the second derivative of $u$ is exactly $f$. In the present case, it could happen that the derivatives $\partial_x^2u$ and $\partial_y^2u$ have singularities that cancel when the sum is performed.

A good way to study the regularity of $u$ is to consider the
Fourier transform $\hat{u}$ of $u$. Indeed, regularity properties of $u$ are encoded in the decay properties at infinity of $\hat{u}$. This can be seen from the following (formal) identity:
\[\nabla u(x)\ =\ (2\pi )^{-1}\int_{\mathbb{R} ^2}e^{ixp}\, ip\hat{u}(p)\, dp\, .\]
If $\hat{u}$ has a fast enough `decay' in $|p|$ at infinity, it makes the above integral absolutely convergent and $u'$ is nice.

After Fourier transformation, the equation becomes $|p|^2\hat{u}(p)=\hat{f}(p)$. For $|p|\geq 1$, we get $\hat{u}(p)=|p|^{-2}\hat{f}(p)$. So the `decay' of $\hat{u}$ is better than that of $\hat{f}$ and $u$ is more regular than $f$. This is called \textit{elliptic regularity}. 

Now we can also apply this method to the equation $\nabla ^2u+Vu=f$ on $\mathbb{R} ^2$, where $V$ is a function. For instance, if $V$ and $f$ are smooth, we can write $\hat{u}(p)=|p|^{-2}\hat{g}(p)$, where $g=f-Vu$. Given some regularity for $u$, we can improve it by the previous formula. The improved regularity can be plugged into the formula again to get a better regularity, and so on. If $V$ or $f$ has a limited regularity, so
does $u$.

Sometimes one is forced to view the equation $\nabla ^2u+Vu=f$ in the distributional sense. In this case, one can still follow the above argument and, when $f$ and $V$ are smooth, so is also $u$, and $u$ satisfies the equation $\nabla ^2u+Vu=f$ in the usual sense.

In the main text, we used the elliptic regularity to get the {\em real analyticity} of the solution of an equation. Roughly speaking, a function $u$ on $\mathbb{R} ^2$ is real analytic if, near any point $(x_0, y_0)$, it can be
written as a polynomial of infinite degree (i.e. a series in powers $(x-x_0)^i(y-y_0)^j$). A real analytic function is always smooth but the function $(x, y)\mapsto xe^{-1/x}$, if $x>0$, and $(x, y)\mapsto 0$, if $x\leq 0$, is actually smooth but not analytic. 

If $V$ and $f$ are analytic, then so is any solution of $\nabla ^2u+Vu=f$. The argument to see this\cite{LH:76} is more involved than the one above. In the Coulomb case, the potential is real analytic away from the set $\Sigma$ of collisions. So the
above arguments apply away from $\Sigma$. If you think about the hydrogen atom and set $r=\sqrt{x^2+y^2+z^2}$, the function $(x, y, z)\mapsto e^{-r/2}$ is a solution of $\nabla ^2u+u(1/r-1/4)=0$. The potential and the solution are real analytic away from $0$ (and the latter has to be since the equation is elliptic) but they both are not even smooth at $0$.


\begin{thebibliography}{199}
\bibitem{fn3} Here `nasty' captures the fact that the function is not differentiable in the usual sense, and not locally integrable near $R=0$.
\bibitem{fn4} This will require information about the behaviour of both $\overline{\chi}$ and $\overline{\phi}$ near $\Sigma_n$ and at $\infty$.
\end{thebibliography}
\end{document}